\documentclass[12pt]{article}

\usepackage{epsf}
\usepackage[dvips]{graphicx}

\begin{document}

\begin{center}
{\bfseries  SEARCH FOR SIGNAL ON PERCOLATION CLUSTER FORMATION
IN NUCLEUS-NUCLEUS COLLISIONS AT RELATIVISTIC ENERGIES}

\vskip 5mm

O.B. Abdinov $^{1}$, A. Kravcakova$^{2}$, A.A. Kuznetsov$^{3}$,
M.K. Suleymanov$^{1,2 \dag}$, A. S. Vodopianov$^{3}$,
S. Vokal$^{2,3}$

\vskip 5mm

{\small
(1) {\it Physics Institute of AS Baku, Azerbaijan}
\\
(2) {\it University of P.J. Safarik, Kosice, Slovakia }
\\
(3) {\it VBLHE, JINR, Dubna, Russia }
\\
$\dag$ {\it E-mail: mais@sunhe.jinr.ru }}
\end{center}

\vskip 5mm

\begin{center}
\begin{minipage}{150mm}
\centerline{\bf Abstract}

The appearance of the strongly interacting matter mixed phase 
(MP)has been suggested to consider to understand qualitatively the regime 
change existence in the behavior of some centrality depending characteristics 
of events. The MP has been 
predicted by QCD for the temperatures around the 
critical temperature Tc and could be formed as a result of nucleon 
percolation in the density nuclear matter. Our main goal is to get a new 
experimental  confirmation  of the  percolation cluster formation as an 
accompanying effect of the MP formation. To reach the goal, the  
experimental data on $Kr+Em$ - reaction  at 0.95 GeV/nuc and 
$Au+Em$ - reaction  at 10.6 GeV/nucl. with a number of target fragments 
$N_h > 8 $, have been analyzed. The behavior of the distributions of the target and the projectile fragments
 have been studied. The experimental data  
have been compared of the data coming from the cascade-evaporation 
model.
We can conclude that:

	   -- the centrality of the collision could be defined  as a 
number of the 
	   target g-fragments in $Kr + Em$ reactions at energies 
	   0.95 A GeV/nucl and as a number of projectile $F$-fragments 
	   with $Z\ge 1$ in $Au + Em$ reactions at energies 
	   10.6 A GeV/nucl; 
          
	  -- the formation of the percolation cluster sufficiently influences 
	  the characteristics of nuclear fragments;          
         
	  -- there are points of the regime changes in the behavior of some 
	  characteristics of $s$-particles as a function of centrality 
	  which could be qualitatively understood as a result of the big 
	  percolation cluster formation.

\end{minipage}
\end{center}

\vskip 10mm

\section{ Introduction }

{\it Mixed Phase}: Studying  the behavior of the hadron-nuclear and
nuclear-nuclear interactions  characteristics as a function of  collision
centrality $Q$ is an important experimental method to get  information
about  changes of the nuclear matter phase, because the increasing $Q$ could
lead to the growth of the nuclear matter baryon density. In other words,
the regime change in the behavior of some centrality depending characteristics
of events is expected by  varying of  $Q$ to be a signal on phase
transition. This method is considered to be the best tool for reaching the
quark-gluon plasma phase of strongly interacting matter. Some experimental
results have already demonstrated the existence of the regime changes
in the event characteristics behavior as a function of the collision
centrality \cite{1}-\cite{8}. The regularity is observed for
hadron-nuclear \cite{1}-\cite{2}, heavy ~\cite{3}-\cite{7} and light
nuclear \cite{8} interactions in a large domain of nuclear masses and
initial energies. It has been also observed for the behavior of
some centrality characteristics of  $\pi$-mesons , nucleons, fragments,
strange particles, and even for those of $J/\psi$. So, the regime 
changes under 
consideration can not be related with the existence of the predicted
QCD point for the hadronic matter quark-gluon phase transition and 
therefore it
has been suggested ~\cite{sssz} to consider the appearance of the strongly 
interacting matter mixed phase (MP) for  qualitative understanding of the regularity. 
MP has been predicted by QCD for the temperatures around the
critical temperature Tc and could be formed as a result of nucleon
percolation in the density nuclear matter. It is related with the 
following.

{\it Percolation cluster}: It is well known that the statistical and
percolation theories can describe critical phenomena best of all and from 
the other hand the regime changes under consideration have  been also observed
for small density and temperature at which the conditions to apply
statistical theories are practically absent. So, one could say that the
percolation approach is practically the only one to describe  the
results.  Paper \cite{9} discusses that percolation clusters are much
larger than hadrons, within which color is not confined; deconfinement
is thus related to the percolation cluster formation. This is the central
topic of the percolation theory, and hence, the connection between percolation
and deconfinement seems to be  very likely \cite{10}. So, the  
experimental information on the particular conditions of the  MP formation 
could be
very important to fix the onset stage of deconfiment for its future
identification. To extract the signals on the accompanied effects
of MP could be one of the ways to get the experimental information on the
MP formation. The percolation cluster formation could be one of these effects
where the MP formation starts at high energies.

{\it Physical picture}: We  can consider the following physical
picture to understand qualitatively the mentioned above.

{\it{\bf At low energies}}: At some critical values of centrality
$Qc$  the compressed compound nuclear
system could appear. In this system the thermal equilibrium could be
established as a result of Fermi motion and the percolation occurs
that would result in big percolation cluster formation which will then 
be
fragment on the nuclear fragments. So the process of the percolation cluster
could  influence  the nuclear fragments characteristics. This idea was
experimentally tested in \cite{11} for high energy interactions. Section
2 shows the experimental results for the heavy nuclear interaction
at the low energies.

{\it {\bf At middle and high energies}}. First we have to note that  in
comparison with the low energy interaction at middle and high energies the
contributions of multiparticle collisions have to increase strongly
(particularly in the region of central collisions near $Qc$) and
the quark-gluon degree of freedom of the matter could appear.  Paper
\cite{12} discusses that the hadron-chemical-equilibrium could be
established as a result of multiparticle collisions during the heavy
nuclear interactions. In this system the percolation could occur and the
big percolation  cluster might be formed. But in comparison with the low 
energy picture in this case the percolation cluster could consist of hadrons 
and quarks representing  a mixed phase. One more issue to be considered
in section 2 is a possibility to get the signal on percolation cluster
formation in  heavy nuclear interactions at high energies.

As we have mentioned above, the idea that the process of the percolation cluster 
could influence the nuclear fragments characteristics was experimentally
tested in \cite{11}. It will be the main idea to get the information on
the percolation cluster formation. To reach this goal,  two ways of $Q$
determination were used in paper \cite{11}. In one way the values of
$Q$ were determined as a number of protons emitted in one event and in the
second one -- as a number of protons and fragments emitted in one event.
The events of $^{12}CC$-interaction at the momentum of  4.2 A GeV/c were
used  \cite{13}. The experimental data were compared with the simulation
data coming from the  quark-gluon string model (QGSM) without the nuclear
fragments \cite{14}.  It is supposed that  the behavior of  the 
events'number  dependent of $Q$ determined the both ways should be  
similar if there are no clusters as a source of fragments and they would 
differ if the cluster exists  as one. It was obtained that  the form of the
distribution strongly differs for the distribution with different $Q$
determination ways. In the second case  the  two steps structure was
indicated  in the behaviour of the  distribution which could not be
described  by the model. This result has demonstrated that the influence
of nuclear fragmentation processes on the behaviour of the events
number dependent of $Q$  has a critical character. But it is clear that
the light nuclear interaction is not a good object to study the fragmentation
processes. The main properties of the nuclear fragmentation were obtained at low and middle
energy collisions of heavy nuclei \cite{15}. So, we turn to the low and high
energy collisions of heavy nuclei and our main goal is to get a new
experimental  confirmation  of the  percolation cluster formation as an
accompanying effect of the MP formation.

{\it Centrality of  the collisions}: Before discussing the experimental
results we would like to touch upon  one more question which  is very
important for the centrality experiments. It is clear that the  centrality
of collisions $Q$ can not be  defined directly in the experiment. In
different   experiments the values of $Q$ are defined as a number of
identified protons , projectiles'  and
 targets' fragments,  slow particles, all particles, as the
 energy flow of the particles with emission  angels $\theta \simeq 0^0$  or
 with  $\theta \simeq 90^0$ . Apparently, it is not simple to compare
 quantitatively the results on $Q$-dependencies obtained in  different
 papers and from the other hand the definition of $Q$ could 
significantly
 influence the final results.  So we believe it is necessary to understand
 what centrality $Q$ is.  Usually for a chosen variable to fix $Q$ it is
 supposed that its values have to increase linearly with a number of colliding
 nucleons or baryon density of the nuclear matter. The simplest mechanism that
 could give this dependence is the cascade approach. So, we have used one 
of the
 versions of the cascade-evaporation model CEM \cite{16} to choose the variable
 to fix $Q$ for studying the centrality dependence of the event characteristics.

\section{Experiment }

{\it Distribution of the fragments}.To reach the goal, we have analyzed the
experimental data on $Kr+Em$ - reaction  at 0.95 GeV/nucl \cite{17} and
$Au+Em$ - reaction  at 10.6 GeV/nucl. \cite{18}. We have considered the
events with a number of $N_h > 8 $ to select the heavy nuclear collisions
(in papers \cite{17}- \cite{18} this condition was not used). According to
the idea mentioned above  on the centrality event selection, we have 
studied the behavior of the distributions of the target   fragments ( $g$- and $h$-
fragments) and the projectile fragments with charge  $Z\ge 1$
($F$-fragments) . The experimental data have been compared of the  
once coming from the CEM ~\cite{16}.

      {\it The $Kr + Em$ reactions at 0.95 GeV/nucl.} Fig 1a-f
      shows the yields of  $g$ - , $h$ - and $F$-fragments  in the
      $Kr + Em$ (at 0.95 GeV/nucl, Fig. 1 a-c.) and in $Au + Em$ (at 10.6 GeV/nucl, Fig. 1d-f.)
      reactions. The results coming from the CEM  are also drawn.  We
      can see that :

       The $g$-fragments experimental multiplicity distribution ($N_g$)
       for the $Kr + Em$ reactions is well described by the model (Fig. 1a).
       We would remind that  in the framework of the CEM $g$-fragments are
       considered as the results of cascading collisions in the target
       spectator and participant. So $N_g$  could be used to fix the
       centrality of collisions (result I).

       The target  $h$-fragments multiplicity ($N_h$) distribution shape for
       the $Kr + Em$ reactions (Fig. 1b) cannot be described by the CEM in
       the region of $N_h$ values $15 < N_h < 32$ . If we remember that the
       $N_h$ is the number of the final state target fragments in the event  which
       is the sum of the target black fragments ($N_b$) and $N_g$,  we would say
       that the CEM can't describe  the multiplicity distributions of
       $b$-particles which are the slowest target fragments and thus they 
have
       to get much more information on the state of the nuclear target. In 
a
       recent paper ~\cite{19}  one of the authors  using CEM has shown 
that
       to describe fully the $b$-particles yields, it is necessary to take into
       account the percolation mechanism and formation  of big percolation
       cluster. So, we could assert that this observed difference between the
       behavior of the experimental and model $N_h$ - distributions, is 
related
       with the formation of the  big percolation cluster (result II);

       The behavior of the experimental distribution of projectile fragments
       with $Z \ge 1$ produced in $Kr + Em$ collisions ( Fig. 1c) is
       not in agreement with the result coming from the CEM in a full area
       of the $N_F$  definition either. At the point $N_F = 2$ the model gives the
       result more than one order higher in comparison with the experiment. Two other
       $N_F$      regions are observed (at $7 < N_F < 14$ and
       30 < $N_F < 40$)  where  the obtained  model and  the
       experimental data do not agree with each other and we can see that
       the deference has a critical character because it appears only at some
       values of $N_F$ (result III).  The formation of the big percolation
       cluster could give this critical behavior, for example, as the result
       of appearance of the physical picture described  above (for low energy
       interactions).

          {\it The $Au + Em$ reactions at 10.6 GeV/nucl.} The experimental
      distribution of $g$-particles from $Au-Em$ reactions ( Fig. 1d)
      can not be described by the model in full region of the $N_g$
      definition. We can separate some region in the relative between
      the behaviors of the experimental and model distributions. In
      the region of  $N_g < 5$ the model can describe the experimental
      distribution. In the region of $N_g > 15$ the experimental
      values of $N_i$ decrease with $N_g$ while the values coming
      from the model are constant  in the region $15< N_g < 40$. The
      model could  not describe the distribution of $h$-particles
      in a full region of the $N_h$-definition either, that is seen from
      Fig. 1e. The model could only describe the experimental
      distribution of $N_h$ in the region of $22< N_h < 32$.
      So, we can say that for the  reaction  under consideration the
      $N_g$  as well as  the $N_g$ could not be used to fix the
      centrality of collisions (result IV). We believe that the result
      could also be understood qualitatively in the framework of the
      above-mentioned physical picture (for high energy interactions).
      In a  recent paper ~\cite{20} the bond percolation model is used to
      interpret 10.2 GeV/c $p + Au$ multifragmentation data. The critical
      value of the percolation parameter $p_c = 0.65$ was found from the
      analysis of the intermediate mass fragments charge distribution.

          The distribution of projectile fragments with $Z \ge 1$ produced
      in $Au + Em$ collisions is  in good  agreement with the result
      coming from the CEM. So, we can see that the projectile fragments
      are produced by the mechanism similar to the cascade-evaporation
      one and $N_F$ could be used to fix the centrality for these reactions
      (result V).

{\it Correlation.}  It is clear that the obtained results are not sufficient
to  confirm fully  the percolation cluster formation especially at high energy
collisions for which the contributions of multiparticle collisions have to
increase strongly, and particularly in the region of central collisions near the
critical values of centrality. As it has been mentioned, at these high 
energies
the hadron-chemica-equilibrum was established and the percolation could
occur. But in comparison with the low energy physical picture, in this case the
percolation cluster could consist of hadrons and quarks and represents  
the mixed phase.

Thus, one needs to get additional information in future to confirm  the
percolation cluster formation.

A number of the final-state-relativistic single charged particles (s) in the
emulsion experiments (it is called the multiplicity of the shower particles and
is denoted by $<n_s>$)  might be most sensitive to the dynamics of the
interaction at high energies ( as well as the values of pseudorapidity
$\eta$ of $s$-particles) . So, we have studied  the correlation between
the characteristics of s-particles and the values of centrality. As we
have mentioned above,  to fix the centrality, one might use the variable $N_g$
for $Kr + Em$ reactions and $N_F$  for $Au + Em$ ones. Here we discuss
the results  of our study.

Fig. 2a-c presents the average values of multiplicity $<n_s>$
for $s$ - particles produced in $Kr + Em$ and  $Au + Em$  reactions and
the average values of pseudorapidity for $s$ - particles produced in
$Au + Em$  reactions . We can say that there are two regions in the
behavior of the  values  of  $<N_s>$ as a function of  $N_g$  for the
$Kr+Em$  reaction ( Fig. 2a). In the region of : $N_g  < 40$ the
values of  $<N_s>$ increase linearly with $N_g$ , here  the  CEM also
gives the linear dependence but with the slope  less than the experimental
one; $N_g  > 40$ the CEM gives the values for $N_s$  greater than  the
experimental observed ones, the last saturates in this region, the effect
could not be described by the CEM . It have been previously  observed in 
emulsion experiments ~\cite{15}. It is clear that there should be some 
effects  which could stop ( or sufficiently moderate) the increase of  $N_s$. The effect of
the percolation cluster formation could be one of those effects. The 
moderation of the values of $N_s$ as a function of $N_F $ is also observed for the $Au+Em$
reaction at 10.6 GeV/nucl. (Fig. 2c) near the point of  the $N_F\simeq 40-50$
these should  be a point of the regime change which is absent for 
the distribution coming from the CEM.

      Thus, we can say that the effects which could stop
      (or sufficiently
      moderate) the dependence of $<N_s>$  as  a function of centrality
      appear at some values of $N_g$ and $N_F$. It  strengthens the
      result VII because the process of percolation cluster formation is
      a critical effect which  appears at some critical values of centrality.
      If we  compare the behavior of the experimental and  theoretical
       distributions for the values of  $\eta$  of s-particles produced in the
       $Au+Em$ reaction as a function of $N_F$ (Fig. 1c), we would get one
       more confirmation on the existence of the point
       $N_F \simeq 40-50$, behind  which the values of  $\eta $ are
       systematically less than the CEM expectation. But in the region of
       $N_F > 40-50$ the model describes the  experimental
       distribution rather well.

So, we can say that the points of the regime change are  observed in the
behavior of the characteristics of $s$- particles as a function of
centrality. In the central collisions region the increase of the average
values of multiplicities  are  sufficiently moderate (or stopped) and the
average values of  $\eta$ decrease and could not be described by the CEM.
It could be qualitatively understood  within the formation of the big percolation cluster.

\section{Conclusion}

          We can conclude that:

       -- the centrality of collision could be defined of as a number of 
the
       target g-fragments in $Kr + Em$ reactions at energies
       0.95 A GeV/nucl and as a number of projectile $F$-fragments
       with $Z\ge 1$ in $Au + Em$ reactions at energies
       10.6 A GeV/nucl;

      -- the formation of the percolation cluster sufficiently influences
      the characteristics of nuclear fragments;

      -- there are points of the regime changes in the behavior of some
      characteristics of $s$-particles as a function of centrality
      which could be qualitatively understood as a result of the big
      percolation cluster formation.

\newpage

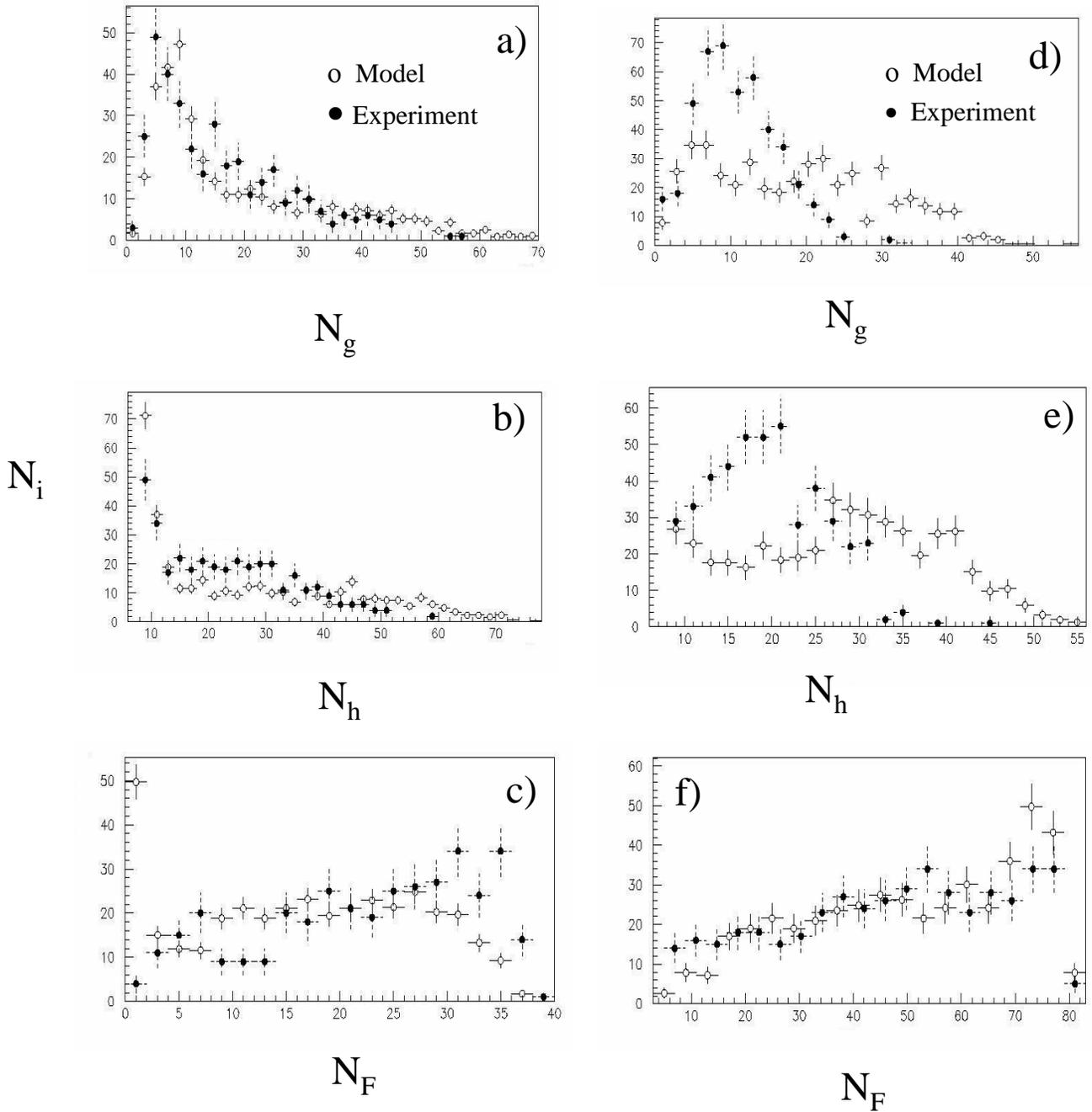
Fig. 1a-f. Distribution of the  target a) $g$ - fragments; b) 
$h$-fragments;  projectiles' c) $F$-fragments produced  
in the $Kr + Em$ reactions at 0.95 GeV/nucl and target d) $g$ - 
fragments; e) $h$-fragments;  projectile 
f) $F$-fragments produced  in the $Au + Em$ reactions at 10.6 GeV/nucl. 
It  also gives the results coming from the CEM calculation.

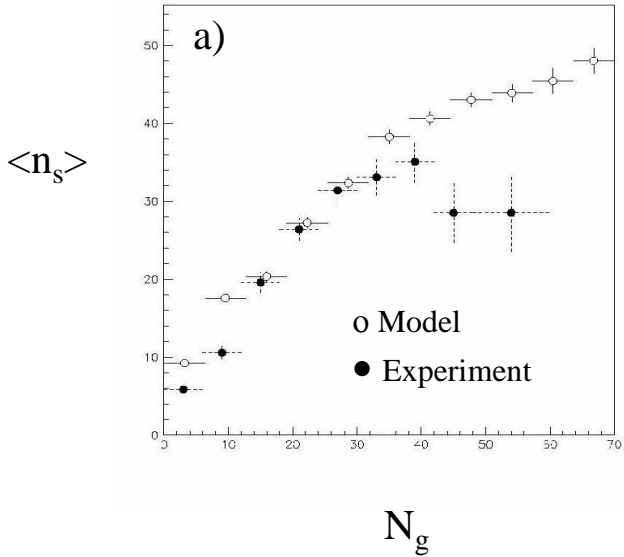
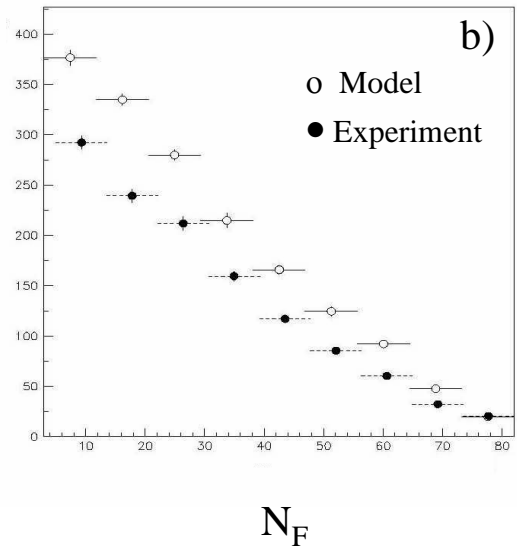
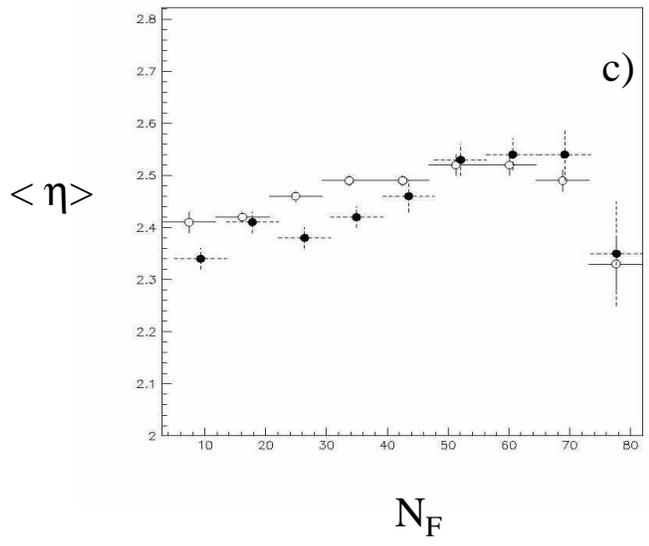
Fig. 2a-c. a) average values of  $s$ - particles multiplicity produced 
 in the $Kr + Em$ reactions at 0.95 GeV/nucl. as a function of  $N_g$ ; 
average values of the b) multiplicity; c) pseudorapidity   of s-particles 
produced in 
$Au + Em$ reactions  at 10.6 GeV/nucl. as a function of  $N_F$ .  
It also shows the result coming from the CEM.

\end{document}